\documentclass[nofootinbib,twocolumn]{an}
\usepackage{amsmath}
\usepackage{graphicx}
\usepackage{times}

\begin{document}

\title{Black hole tidal charge constrained by strong gravitational lensing}
\author{ Zs. Horv\'{a}th \inst{1}\thanks{\email{zshorvath@titan.physx.u-szeged.hu}}\and L.A. Gergely \inst{1}}

\authorrunning{Zs. Horv\'{a}th \& L. A. Gergely}
\titlerunning{Black hole tidal charge constrained...}

\institute{
$^{1}$ Departments of Theoretical and Experimental Physics, University of Szeged, D\'{o}m t\'{e}r 9, Szeged 6720, Hungary}


\abstract{Spherically symmetric brane black holes have tidal charge, which modifies
both weak and strong lensing characteristics. Even if lensing measurements
are in agreement with a Schwarzschild lens, the margin of error of the
detecting instrument allows for a certain tidal charge. In this paper we
derive the respective constraint on the tidal charge of the supermassive
black hole (SMBH) in the center of our galaxy, from the radius of the first
relativistic Einstein ring, emerging in strong lensing. We find that even if
general relativistic predictions are confirmed by high precision strong
lensing measurements, SMBHs could have a much larger tidal charge, than the
Sun or neutron stars.}
\maketitle
\section{Introduction}

The Galactic Center is hard to resolve due to source confusion. Absorbing
gas and dust renders most of the knowledge about this part of our galaxy to
come from observations in radio and infrared (Ghez et al. 2003)\nocite{Ghez1}%
. It has been estimated (Brown et al. 2005) \nocite{r5}that cold dark matter
remnants of $1000$ solar masses (M$_{\odot }$) and compact star remnants of $%
1000$ M$_{\odot }$ reside in the inner $0.01$ parsec (pc) region. The star
population in the inner $0.04$ pc, the Sgr A* stellar cluster consists of B
stars (Krabbe et al. 1995)\nocite{r3}. In the distance range [0.04 pc, 0.5
pc] Wolf Rayet and OB giant stars are found.

The Galactic Center is dominated by a Supermassive Black Hole (SMBH), with
mass $4.31\times 10^{6}$ M$_{\odot }$ and distance from the Sun $8330$ pc
(Gillessen et al. 2012)\nocite{Gillessen2}. The orbits of nearby stars
depend on the mass and spin of the SMBH. These characteristics were obtained
observing of stellar orbits (Will 2008; Merritt et al. 2010).\nocite{will}%
\nocite{Merritt}

The first measurements of proper motions of stars within $2400$ AU from the
center of our Galaxy was published by Eckart and Genzel (1997)\nocite{ECKART}%
. A detailed review of 16 years of monitoring stellar orbits around the SMBH
using near-infrared (NIR) techniques was given by Gillessen et al. (2009)%
\nocite{Gillessen2009}.

With increasing measurement accuracy and the possibility of direct
observations of the horizon of the SMBH at the Galactic Center the
possibility of testing general relativistic predictions emerges. The
measurement accuracy of a specific instrument will allow for certain margin
of error for any additional parameter.

The design of the interferometer GRAVITY has reached completion (Gillessen
et al. 2010)\nocite{Gillessen}. This will use the four telescopes of the
Very Large Telescope as an interferometer in the NIR band. GRAVITY will
perform astrometry with $12$ microarcsecond ($\mathrm{\mu a}$\textrm{s})
precision in the K band up to 15 magnitudes. The margin of error of its
measurements will be $\Delta \Theta =12\times 10^{-6}~\mathrm{as.}$ From the
numerous scientific applications of GRAVITY (Gillessen et al. 2010, Section
3. )\nocite{Gillessen}, the observation of relativistic motions near the
horizon of Sgr A* will be of uttermost importance.

By such observations of NIR flares the metric near the horizon can be
determined (Psaltis 2004)\nocite{pp}. A submm-VLBI array should be able to
actually resolve Sgr A* (Falcke, Melia \& Agol 2000).\nocite{f} As proposed
by Will (2008)\nocite{will} the no-hair theorem can be tested.

We will constrain the tidal charge, an imprint of the possible 5-dimensional
character of gravity in a brane-world scenario (Maartens 2000; Maartens \&
Koyama 2010). \nocite{MaartensPRD}\nocite{MaartensKoyamaLivRev} The tidal
charge emerges from the Weyl curvature of a 5-dimen\- sional space-time in
which our 4-dimensional observable world is embedded. Such a tidal charged
black hole solution on a brane was found by Dadhich et al. (2000)\nocite%
{dmpr}. In these theories the electric part of the Weyl curvature,
non-standard model 5-dimensional fields, asymmetric embedding and a varying
brane tension all could act as sources of the effective Einstein equation
(Shiromizu, Maeda \& Sasaki 2000; Gergely 2003, 2008).\nocite{SMS}\nocite%
{Decomp}\nocite{VarBraneTension} Remarkably, the Weyl curvature source term
could replace dark matter (Harko \& Cheng 2007; Mak \& Harko 2004) \nocite%
{HarkoRC}\nocite{HarkoClusters}.

Light deflection and weak gravitational lensing by tidal charged brane black
holes was investigated by B\"{o}hmer, Har- ko and Lobo (2008) ; Gergely,
Keresztes \& Dwornik (2009).\nocite{HarkoLens}\nocite{GergelyLens} One of
the results of these investigations was a strict limit imposed on the tidal
charge from Solar System measurements. Horv\'{a}th, Gergely and Hobill
(2010) \nocite{HorvathLens}derived analytical results on weak lensing by
tidal charge dominated black holes. Bin-Nun (2010b) \nocite{BinNun2}
discussed weak lensing by the Galactic SMBH assumed tidal charged. The
angular position of the image and its apparent magnitude were shown to both
decrease as function of the tidal charge.

Whisker (2005)\nocite{Whisker} discussed strong lensing in two brane-world
black hole models, the "$U=0$ black hole"\ and the tidal charged black hole.
For the central black hole of our galaxy as lens the angular radius where
the second and higher relativistic images are confused was given. The
angular separation of these relativistic images from the first relativistic
image was also presented. Bin-Nun (2010a)\nocite{BinNun} has discussed three
brane BH solutions, including the tidal charged one. The size and
magnifications of the primary and the first two relativistic Einstein rings
was given for a source lensed by the SMBH at Sgr A*. Bin-Nun (2011)\nocite%
{BinNun3} extends the analysis presented by Bin-Nun (2010b)\nocite{BinNun2}
to the first relativistic images of the stars S2, S6, S14.

In this paper we study strongly relativistic orbits in the Galactic Center
assuming that the metric is the one of the tidal charged brane black hole.
We focus on 1-loop photon trajectories, as depicted on Fig. 1. We summarize
the basic equations concerning the one-loop null geodesics in spherically
symmetric, static space-times in Section \ref{geod}. The formation of the
first relativistic Einstein ring, emerging due to strong lensing by the SMBH
in the center of our galaxy is discussed in Section \ref{gyuru}. Then we
investigate the magnitude of the tidal charge, still compatible with the
margin of error established by GRAVITY for the radius of the first
relativistic Einstein ring.

We present our conclusions which include the constraints on the tidal charge
derived by this method in Section \ref{Concl}. We use the geometric units, $%
G=c=1$. 
\begin{figure*}[tbp]
\includegraphics[width=15cm]{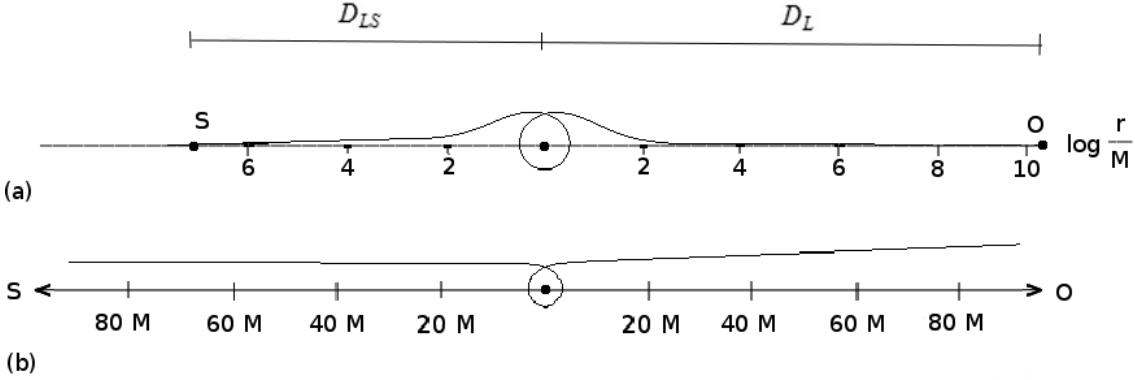}
\caption{The trajectory of light along which the photons travel from the
source $\mathrm{S}$ to the observer $\mathrm{O}$, while turning around the
lens $\mathrm{L}$ once, for $q=0$, $D_{\mathrm{L}}=8600$ pc$,$ $D_{\mathrm{LS%
}}=10$ pc. On panel (a) the whole geodesics is seen on logarithmic scale.
The distorted region in the left and right side of the loop is magnified and
shown on a linear scale on panel (b). }
\label{Fig1}
\end{figure*}

\section{Null geodesics and relativistic Einstein rings in spherically
symmetric, static space-times \label{geod}}

The general spherically symmetric, static metric is 
\begin{equation}
ds^{2}=g_{tt}dt^{2}+g_{rr}dr^{2}+r^{2}d\theta ^{2}+r^{2}sin^{2}\theta {\ }%
d\varphi ^{2}{~.}  \label{gsz}
\end{equation}%
In the following we consider metrics with $g_{\theta \theta }=r^{2}$, $%
g_{\varphi \varphi }=r^{2}sin^{2}\theta .$ Without reducing generality, we
can restrict geodesic motions to the plane $\theta =\pi /2$. We also
introduce the dimensionless radial coordinate $R=r/M$. The second order
radial geodesic equation can be replaced by the first order null condition $%
0=g_{tt}\left( dt/dp\right) ^{2}+g_{rr}\left( dr/dp\right) ^{2}$ $%
+g_{\varphi \varphi }\left( d\varphi /dp\right) ^{2}$ where $p$ is a
parameter of the null curve). As the metric does not depend on either of the
coordinates $t$ or $\varphi $, two constants of motion emerge by integration
from the geodesic equations: $L=g_{\varphi \varphi }d\varphi /dp$ the
specific angular momentum of the photon with dimension [length$^{2}]$; and $%
E=g_{tt}dt/dp$, proportional to the energy of the photon, the dimension of $E
$ being [length].

Then the $rr$ term can be written in the form $g_{rr}\left( dr/d\varphi
\right) ^{2}$ $\times L^{2}/g_{\varphi \varphi }^{2}$, while the $tt$ and $%
\varphi \varphi $ terms as $E^{2}/g_{tt}$ and $L^{2}/g_{\varphi \varphi }$.
Hence the null condition becomes 
\begin{equation}
0=\frac{E^{2}}{g_{tt}}+g_{rr}\left( \frac{dr}{d\varphi }\right) ^{2} \frac{%
L^{2}}{g_{\varphi \varphi }^{2}}+\frac{L^{2}}{g_{\varphi \varphi }} {~.}
\end{equation}
By reordering the terms, we obtain the equation characterizing the
trajectory $r\left( \varphi \right) $: 
\begin{equation}
\frac{dr}{d\varphi }=\pm \left[ \frac{g_{\varphi \varphi }}{g_{rr}}\left( 
\frac{E^{2}}{L^{2}}\frac{g_{\varphi \varphi }}{-g_{tt}}-1\right) \right]
^{1/2}{~.}  \label{PalyaAlak}
\end{equation}%
The sign differentiates between the incoming and outgoing parts of the path.

We will seek a null geodesic curve along which the photons travel from the
source $\mathrm{S}$ to the observer $\mathrm{O}$, while turning around the
lens $\mathrm{L}$\textrm{\ }once (1-loop orbit). The source, the lens and
the observer are placed on the same coordinate line $\varphi =0$ (the
optical axis). More specifically $\mathrm{L}$ is the origin, $\mathrm{S}$
lies at $\left( \varphi =0,r=D_{\mathrm{LS}}\right) $, while $O$ at $\left(
\varphi =\pi ,r=D_{\mathrm{L}}\right) $. Then the trajectory $r\left(
\varphi \right) $ of the photon is a monotonically decreasing function from $%
r=D_{\mathrm{LS}}$ to some distance $r=r_{\mathrm{\mathrm{min}}}$, then
increases to $r=D_{\mathrm{L}}$. The distance of minimal approach $r_{%
\mathrm{\mathrm{min}}}=r\left( \varphi _{\mathrm{min}}\right) $ is defined
by the equation 
\begin{equation}
\frac{dr}{d\varphi }\left( \varphi _{\mathrm{min}}\right) =0~.
\label{rminDEF}
\end{equation}%
For such a one-loop path the total change of the polar angle $\varphi \,$\
from $\mathrm{S}$ to $\mathrm{O}$ is $\pi +2\pi $.

The image created by these 1-loop orbits is called the first relativistic
Einstein ring. Its angular radius, the first relativistic Einstein angle $%
\Theta _{\mathrm{E}}$ is defined as the angle between the optical axis $%
\mathrm{SO}$ and the tangent $\partial /\partial p$ of the geodesic curve $%
R\left( \varphi \right) $ at the point $\mathrm{O},$ where $R_{\mathrm{O}%
}=D_{\mathrm{L}}/M$. The scalar product of the vector $\partial /\partial R$
(giving the direction of the optical axis $\mathrm{SO}$ at $\mathrm{O}$)
with the tangent $\partial /\partial p$ can be evaluated at $R_{\mathrm{O}}$
two ways. First, from the definition of the scalar product:%
\[
\frac{\partial }{\partial R}\cdot \frac{\partial }{\partial p}=\left\vert 
\frac{\partial }{\partial R}\right\vert \left\vert \frac{\partial }{\partial
p}\right\vert \cos \Theta _{\mathrm{E}}= 
\]%
\[
\sqrt{g_{RR}}\sqrt{\left( \frac{\partial R}{\partial p}\right)
^{2}g_{RR}+\left( \frac{\partial \varphi }{\partial p}\right) ^{2}g_{\varphi
\varphi }}\cos \Theta _{\mathrm{E}}~. 
\]%
Second, from the decomposition of the tangent vector in the polar coordinate
system we get:%
\begin{eqnarray*}
\frac{\partial }{\partial R}\cdot \frac{\partial }{\partial p} &=&\frac{%
\partial }{\partial R}\cdot \left( \frac{\partial R}{\partial p}\frac{%
\partial }{\partial R}+\frac{\partial \varphi }{\partial p}\frac{\partial }{%
\partial \varphi }\right) \\
&=&\frac{\partial R}{\partial p}\frac{\partial }{\partial R}\cdot \frac{%
\partial }{\partial R}=\frac{\partial R}{\partial p}g_{RR}~.
\end{eqnarray*}%
The equality of the right hand sides gives%
\[
\Theta _{\mathrm{E}}\equiv \mathrm{arccos}\frac{\frac{\partial R}{\partial p}%
\sqrt{g_{RR}}}{\sqrt{\left( \frac{\partial R}{\partial p}\right)
^{2}g_{RR}+\left( \frac{\partial \varphi }{\partial p}\right) ^{2}g_{\varphi
\varphi }}} 
\]%
\begin{equation}
=\mathrm{arccos}\left[ \frac{\left( \frac{dR}{d\varphi }\right) ^{2}g_{RR}}{%
\left( \frac{dR}{d\varphi }\right) ^{2}g_{RR}+g_{\varphi \varphi }}\right]
^{1/2}{.\ }  \label{Theta}
\end{equation}%
In the last step the parametrization of the curve $R\left( \varphi \right) $
was eliminated by multiplying both the numerator and the denominator by $%
dp/d\varphi .\ $

\section{The first relativistic Einstein ring and constraints on the tidal
charge \label{gyuru}}

\begin{table}[tbp]
\caption{The allowed range of the Einstein ring (column 1.) and the
associated range of the tidal charge (columns 5.-6.). Columns 2.-3.: the
orbital parameters of the orbits with a polar angle change $3\protect\pi ,$
column 2.: the minimal distance $R_{\mathrm{min}}$, larger than the horizon
in each case, column 3.: the parameter $L/ME$, column 4.: the normalized
radius of the horizon, column 5.: the tidal charge, column 6.: the
normalized tidal charge. We chose $D_{\mathrm{LS}}=10$ pc.}
\label{Table1}%
\begin{tabular}{cccccc}
\hline
$\Theta _{\mathrm{E}}$ & $\frac{r_\mathrm{min}}{M}$ & $\frac{L}{ME}$ & $%
\frac{r_{\mathrm{H}}}{M}$ & $q $ & $\frac{q}{M^{2}}$ \\ 
$\mathrm{\mu as}$ &  &  &  & $\mathrm{\left[ 10^{20}~m^{2}\right] }$ &  \\ 
\hline
38 & 4.9 & 7.7 & 3.3 & -1.815 & -4.5 \\ 
34 & 4.2 & 6.7 & 2.8 & -1.008 & -2.5 \\ 
30 & 3.7 & 6.1 & 2.4 & -0.491 & -1.2 \\ 
26 & 3.0 & 5.2 & 2.0 & 0.000 & 0.0 \\ 
22 & 2.5 & 4.4 & 1.6 & 0.148 & 0.3 \\ 
18 & 1.9 & 3.8 & 1.1 & 0.268 & 0.7 \\ 
14 & 1.4 & 3.2 & n. s. & 0.524 & 1.3 \\ \hline
\end{tabular}%
\end{table}
The tidal charged black hole (Dadhich et al. 2000)\nocite{dmpr} generalizes
the Schwarzschild solution of a black hole with mass $M$ by allowing for a
tidal charge parameter$\ q$ in the metric function 
\begin{equation}
g_{tt}=-\frac{1}{g_{rr}}=-1+\frac{2M}{r}-\frac{q}{r^{2}}~,
\label{TidalMetric}
\end{equation}%
of the spherically symmetric, static space-time (\ref{gsz}). For $q<0$ there
is a horizon at $M+(M^{2}-q)^{1/2}$, while for $0<q<M^{2}$ there are two
horizons, at $M\pm (M^{2}-q)^{1/2}$. For $q>M^{2}$ the metric describes a
naked singularity.

The trajectory (\ref{PalyaAlak}) becomes 
\begin{equation}
\frac{dR}{d\varphi }=\pm \left[ \left( \frac{EM}{L}\right)
^{2}R^{4}-R^{2}+2R-\frac{q}{M^{2}}\right] ^{1/2}{~,\ }  \label{NormaltPalya}
\end{equation}
$L/ME$ denotes the dimensionless impact parameter. The minimal dimensionless
distance $R_{\mathrm{min}}:=r_{\mathrm{min}}/M$ is a root of the polynomial
obtained from Eq. (\ref{NormaltPalya}) by setting the left hand side as 0.

The Einstein angle (\ref{Theta}) then becomes%
\begin{equation}
\Theta _{\mathrm{E}}=\mathrm{arccos}\left[ 1\!-\!\left( \frac{L}{ED_{L}}%
\right) ^{2}\left( \!1-\!\frac{2M}{D_{L}}\!+\!\frac{q}{D_{L}^{2}}\right) %
\right] ~.  \label{einang}
\end{equation}

Next we assume a stellar source on the optical axis defined by the SMBH and
the GALAXY detector, at a distance $D_{\mathrm{LS}}$ varying in the range [$%
10$ pc, $10^{5}$ pc] and a photon trajectory with one loop about the SMBH.
Each value of $L/ME$ determines a null geodesic curve by Eq. (\ref%
{NormaltPalya}), however for most of the values the resulting curve does not
reproduce the desired lensing geometry (e. g. the boundary conditions set by 
$\mathrm{S}$ and $\mathrm{O}$). Therefore one has to `fine-tune` $L/ME$ to
generate the specific curve which satisfies the conditions preset by the
1-loop orbit. We did this in the following way. After setting the tidal
charge $q$ and the distance $D_{\mathrm{LS}}$, we chose an initial value for
the $L/ME$ (between $1$ and $10$). Then we evolved the differential equation
(\ref{NormaltPalya}) numerically and calculated the change in the polar
angle $\varphi $ while the photon travels from $D_{\mathrm{LS}}/M$ to $R_{%
\mathrm{min}}$~then further to $D_{\mathrm{L}}/M$. For $L/ME=10$ this
angular change was too large ($\Delta \varphi >3\pi $) for any $q\in \left[
-5M^{2},{\ }+2M^{2}\right] $ and $D_{\mathrm{LS}}\in $[$10$ pc, $10^{5}$
pc]. Therefore we reduced its value and repeated the procedure, until it
yielded $\Delta \varphi =3\pi $ with $10^{-12}$ rad accuracy. In this
iterative way we obtained the dimensionless impact parameter. Then we
inserted this into Eq. (\ref{einang}) to calculate the radius of the first
relativistic Einstein ring.

Limits $q_{\mathrm{min}}$ and $q_{\mathrm{max}}$ for the tidal charge can be
derived from the designed margin of error of the measurements of GRAVITY.
The values of $\Theta _{\mathrm{E}}$ are enlisted for $D_{\mathrm{LS}}$ $=10$
pc in Table \ref{Table1}. The tidal charge was varied to fit these limits,
the Einstein radius changes within $2\Delta \Theta $. The limits $q_{\mathrm{%
min}}$ and $q_{\mathrm{max}}$ are defined by%
\begin{eqnarray}
\Theta _{\mathrm{E}}(q_{\mathrm{min}},D_{\mathrm{LS}}) &=&\Theta _{\mathrm{E}%
}(0,D_{\mathrm{LS}})+\Delta \Theta ~,  \nonumber \\
\Theta _{\mathrm{E}}(q_{\mathrm{max}},D_{\mathrm{LS}}) &=&\Theta _{\mathrm{E}%
}(0,D_{\mathrm{LS}})-\Delta \Theta ~.  \label{dqdef}
\end{eqnarray}%
We have checked that by varying $D_{\mathrm{LS}}$ in the domain [$10$ pc, $%
10^{5}$ pc] the $\Theta _{\mathrm{E}}$ change by less than 1 $\mathrm{\mu as}
$. With the mass of the SMBH a tidal charge falling in the range [-1.815,
0.524] $\times 10^{20}$ m$^{2}~$\ is allowed. This range is consistent with
measurements of the first relativistic Einstein ring generated by sources
opposite to us with respect to the central SMBH.

\section{Concluding Remarks\label{Concl}}

\begin{table}[tbp]
\caption{Bounds on the tidal charge normalized by mass square. Column 2.:
from observations in the Solar System (B\"{o}hmer et al. 2008), column 3.:
from orbital models of high-frequency quasiperiodic oscillations observed in
neutron star binary systems (Kotrlova, Stuchlik \& T\"{o}r\"{o}k 2008),
column 4.: from the constraints on brane tension and the compactness limit
of neutron stars (Germani \& Maartens 2001), column 5.: from forthcoming
strong lensing observations on the Galactic SMBH derived in this paper.}
\label{Table2}%
\begin{tabular}{ccccc}
\hline
Object & Sun & N.s. Binary & Neutron Star & SMBH \\ \hline
$\mathrm{|q/M^{2}|_{max}}$ & 0.003 & 2.339 & 227.647 & 4.485 \\ \hline
\end{tabular}%
\end{table}
Bounds on the tidal charge of various astrophysical objects were derived
earlier in the literature. For neutron stars a limit of $|q|<10^{7}$ m$^{2}$
was established by Kotrlova et al. (2008) \nocite{kotrlova} from orbital
models of high-frequency quasiperiodic oscillations observed in neutron star
binary systems. From the constraint on the brane tension (Germani \&
Maartens 2001, Eq. (30) )\nocite{GermaniMaartens} a weaker limit for \textit{%
negative} tidal charges emerges in the following way. The junction condition
of the uniform density star and its exterior represented by the tidal
charged metric associates a tidal charge to the brane tension, provided the
junction radius is known. The latter is bounded from below by the
compactness limit. This gives $-9.\ 730\times 10^{8}$ m$^{2}<q<0$. The
strongest constraint for the Sun was found by B\"{o}hmer et al. (2008)\nocite%
{HarkoLens} from the perihelion precession of the Earth, $|q|\leq 6\times
10^{3}$ m$^{2}$. Light deflection measurements imposed a milder restriction
on the tidal charge of the Sun, $|q|\leq 2.\ 966\times 10^{9}$ m$^{2}$
(Gergely et al. 2009)\nocite{GergelyLens}.

The question comes, whether experiments available in the near future
targeting the observation of much larger objects in the Universe, where
strong gravitational lensing co- uld be relevant, would lead to other
limits. Such tests concerning relativistic Einstein rings will be achievable
by measurements of the designed instrument GRAVITY. We have studied the
formation of the first relativistic Einstein ring in the tidal charged black
hole geometry. For this we have specified the lensing geometry for the SMBH\
in the center of our galaxy, considered as a lens and for a light source
opposite to us with respect to this SMBH, as a source. Following B\"{o}hmer
et al. (2008), Gergely et al. (2009)\nocite{HarkoLens}\nocite{GergelyLens}
thus assuming that strong lensing experiments will confirm the general
relativistic predictions, but measurement error will still allow for some
tidal charge, we have found the value for the possible tidal charge of the
SMBH in the Center of the Galaxy, which is much larger than the one allowed
for other astrophysical objects. Despite GRAVITY possibly confirming the
Schwarzschild geometry within measurement accuracy, the range [-1. 815, 0.
524] $\times 10^{20}$ m$^{2}$ of the tidal charge would be still allowed.

We have additionally checked that the second and third relativistic rings
(with $\Delta \varphi =5\pi ,7\pi $, respectively) lead to identical results
on the allowed range of $q$, as the first ring does, supporting that the
relativistic rings are situated quite close to each other (within $0.03$ $%
\mathrm{\mu a}$\textrm{s}).

Although the derived constraint on $q$ is much weaker than those from
neutron stars or Solar System observations, the dimensionless quantity $%
q/M^{2}$ is not very much different, as shown in Table \ref{Table2}. In fact
for this quantity the neutron stars constraints are the weakest, and the
Solar System constraints the strongest, while both the neutron star binary
systems and the SMBH strong lensing considerations presented in this paper
gave comparable constraints.

In summary even if general relativistic predictions on the Galactic SMBH are
confirmed by high precision measurements, our investigations show that SMBHs
could have a much larger tidal charge, than the Sun or neutron stars.

\acknowledgements ZsH was supported by the European Union and co-funded by
the European Social Fund through Grant No. T\'{A}MOP 4. 2.
2/B-10/1-2010-0012. L\'{A}G was partially supported by COST Action MP0905
"Black Holes in a Violent Universe".

\end{document}